\documentclass[amsfonts, amssymb, amsmath, floatfix, superscriptaddress, showpacs, aps, reprint]{revtex4-1}
\usepackage{bm}
\usepackage{graphicx}
\usepackage{epstopdf}

\begin{document}
\title{Kondo lattice heavy fermion behavior in CeRh$_2$Ga$_2$}

\author{V. K. Anand}
\altaffiliation{vivekkranand@gmail.com}
\affiliation{\mbox{Helmholtz-Zentrum Berlin f\"{u}r Materialien und Energie GmbH, Hahn-Meitner Platz 1, D-14109 Berlin, Germany}}
\affiliation{ISIS Facility, Rutherford Appleton Laboratory, Chilton, Didcot, Oxon, OX11 0QX, United Kingdom}
\author{D. T. Adroja}
\altaffiliation{devashibhai.adroja@stfc.ac.uk}
\affiliation{ISIS Facility, Rutherford Appleton Laboratory, Chilton, Didcot, Oxon, OX11 0QX, United Kingdom}
\affiliation{\mbox{Highly Correlated Matter Research Group, Physics Department, University of Johannesburg, P.O. Box 524,} Auckland Park 2006, South Africa}
\author{A. Bhattacharyya}
\affiliation{ISIS Facility, Rutherford Appleton Laboratory, Chilton, Didcot, Oxon, OX11 0QX, United Kingdom}
\affiliation{\mbox{Highly Correlated Matter Research Group, Physics Department, University of Johannesburg, P.O. Box 524,} Auckland Park 2006, South Africa}
\author{B. Klemke}
\author{B. Lake}
\affiliation{\mbox{Helmholtz-Zentrum Berlin f\"{u}r Materialien und Energie GmbH, Hahn-Meitner Platz 1, D-14109 Berlin, Germany}}

\date{\today}

\begin{abstract}
The physical properties of an intermetallic compound CeRh$_2$Ga$_2$ have been investigated by magnetic susceptibility $\chi(T)$, isothermal magnetization $M(H)$, heat capacity $C_{\rm p}(T)$, electrical resistivity $\rho(T)$, thermal conductivity $\kappa(T)$ and thermopower $S(T)$ measurements. CeRh$_2$Ga$_2$ is found to crystallize with CaBe$_2$Ge$_2$-type primitive tetragonal structure (space group $P4/nmm$). No evidence of long range magnetic order is seen down to 1.8~K\@. The $\chi(T)$ data show paramagnetic behavior with an effective moment $\mu_{\rm eff} \approx 2.5\,\mu_{\rm B}$/Ce indicating Ce$^{3+}$ valence state of Ce ions. The $\rho(T)$ data exhibit Kondo lattice behavior with a metallic ground state. The low-$T$ $C_{\rm p}(T)$ data yield an enhanced Sommerfeld coefficient $\gamma = 130(2)$~mJ/mol\,K$^2$ characterizing CeRh$_2$Ga$_2$ as a moderate heavy fermion system. The high-$T$ $C_{\rm p}(T)$ and $\rho(T)$ show an anomaly near 255~K, reflecting a phase transition. The $\kappa(T)$ suggests phonon dominated thermal transport with considerably higher values of Lorenz number $L(T)$ compared to the theoretical Sommerfeld value $L_0$. 
\end{abstract}

\maketitle

\section{INTRODUCTION}

Ce-based 122 intermetallic compounds are well known for their diverse magnetic states due to competing Ruderman-Kittel-Kasuya-Yosida (RKKY) and Kondo interactions \cite{Stewart1984,Riseborough, Amato, Lohneysen, Pfleiderer2009, Si2010}. 
The Kondo lattice antiferromagnetic heavy fermion superconductor CeCu$_2$Si$_2$ is one of the most interesting Ce-122 compounds \cite{Steglich, Steglich1996}. Superconductivity in this compound not only coexists with magnetic order but is also believed to be driven by magnetic fluctuations \cite{Stockert2011}. Evidence for multiband superconductivity has also been seen from heat capacity measurements on CeCu$_2$Si$_2$ \cite{Kittaka2014}. Because of the instability of the $4f$ shell of Ce the electronic ground state of Ce-compounds can be tuned by pressure, field or chemical composition. Of particular interest are the systems situated close to a quantum critical point where quantum fluctuations dominate the physics.  The Kondo lattice heavy fermion antiferromagnet CePd$_{2}$Si$_{2}$ is one such interesting compound where the application of pressure tunes the ground state, leading to pressure induced superconductivity and a non-Fermi liquid behavior after the antiferromagnetism is suppressed completely \cite{Grier, vanDijk, Grosche, Demuer}.

Because of the exotic physical properties of Ce-compounds and simple crystal structure of 122 compounds the Ce$T_{2}X_{2}$ ($T$= transition element, $X$ = Ge, Si) compounds have been intensively investigated. While most of the Ce$T_{2}X_{2}$ compounds have been found to crystallize in centrosymmetric ThCr$_{2}$Si$_{2}$-type body-centered tetragonal structure (space group $I4/mmm$), there are a few compounds that form in a noncentrosymmetric variant CaBe$_{2}$Ge$_{2}$-type primitive tetragonal structure (space group $P4/nmm$). Even a polymorphism of the two structures has been seen.   For example, valence fluctuating system CeIr$_{2}$Si$_{2}$ shows polymorphism.  CeIr$_{2}$Si$_{2}$ is found to form in ThCr$_{2}$Si$_{2}$-type body-centered tetragonal structure at low temperature which exhibits Fermi liquid behavior, whereas at high temperature it forms in CaBe$_{2}$Ge$_{2}$-type primitive tetragonal structure that exhibits non-Fermi liquid behavior \cite{Hiebl, Mihalik}.

The Kondo effect (Kondo temperature) strongly depends on the strength of hybridization between the conduction and $4f$ electrons (c-$f$ hybridization), the strength of this c-$f$ hybridization depends on the unit cell volume. Thus a small change in unit cell volume can significantly modify the physical properties of Ce-compounds. The change in c-$f$ hybridization strength brought by the change in unit cell volume on account of a change in crystal structure can thus be one possible factor for the different ground state properties of the polymorphic phases of CeIr$_{2}$Si$_{2}$. CeIr$_{2}$Ge$_{2}$ is another compound that forms in CaBe$_{2}$Ge$_{2}$-type primitive tetragonal structure and exhibits Kondo lattice heavy fermion behavior \cite{Mathur,Sampatkumaran1996, Mallik}. CePd$_{2}$Ga$_{2}$ has a CaBe$_{2}$Ge$_{2}$-type primitive tetragonal structure and undergoes a structural transition to a triclinic phase at 125~K \cite{Kitagawa1999}. CeIr$_{2}$B$_{2}$ forms in a completely different orthorhombic structure (space group \textit{fddd}), and exhibits ferromagnetic ordering  at $T_{\rm c} = 5.1$~K with a weak Kondo interaction \cite{Sampatkumaran,Prasad2012}. 

In our efforts to search for novel Ce-122 compounds we have been successful in synthesizing a new ternary intermetallic compound, namely CeRh$_2$Ga$_2$, which is found to crystallize in CaBe$_2$Ge$_2$-type primitive tetragonal structure. Here we present results of our investigations of physical properties of CeRh$_2$Ga$_2$ based on the magnetic susceptibility $\chi(T)$, isothermal magnetization $M(H)$, heat capacity $C_{\rm p}(T)$, electrical resistivity $\rho(T)$, thermal conductivity $\kappa(T)$ and thermopower $S(T)$ measurements. We find evidence for Kondo lattice behavior with an enhanced Sommerfeld coefficient ($\gamma = 130(2)$~mJ/mol\,K$^2$) revealing a moderate heavy fermion behavior in this compound. There is no clear sign of long range magnetic order down to 1.8~K\@. 

\section{Experimental}

A polycrystalline sample of CeRh$_2$Ga$_2$ was prepared by the standard arc melting technique using stoichiometric amounts of high purity elements. For homogenization the sample was flipped and melted several times during the arc melting process. The weight loss during arc melting was less than 1.0\%. The arc melted sample was further annealed for three weeks at 800~$^{\circ}$C. The crystal structure was determined by powder x-ray diffraction (XRD) using Cu K$_{\alpha}$ radiation. Magnetic measurements [$M$ versus temperature $T$ and $M$ versus magnetic field $H$] were made using Quantum Design magnetic properties measurement system (MPMS) superconducting quantum interference device (SQUID) magnetometer. The $C_{\rm p}(T)$ was measured by adiabatic relaxation technique using the heat capacity option of Quantum Design physical properties measurement system (PPMS). The $\rho(T)$ was measured by standard four probe method using the resistivity option of PPMS. The $\kappa(T)$ and $S(T)$ were measured using the thermal transport option of PPMS. The bulk properties measurements at Helmholtz Zentrum Berlin were made using the MPMS and PPMS facilities of Laboratory for Magnetic Measurements (LaMMB).

\section{\label{Sec:Results} Results and Discussion}

\begin{figure}
\begin{center}
\includegraphics[width=8.5cm]{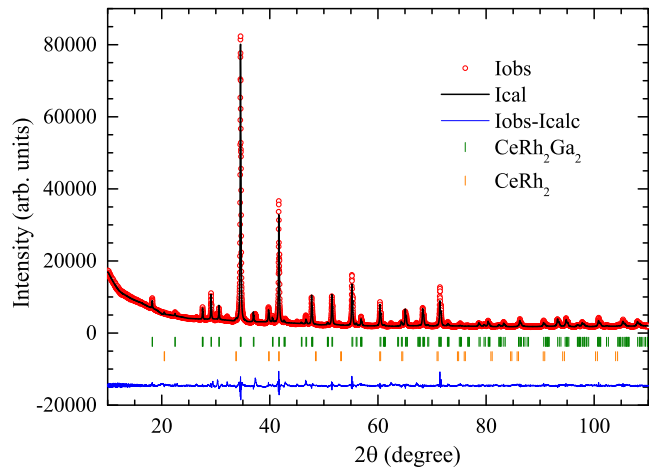}
\caption{\label{fig:XRD} (Color online) Room temperature powder x-ray diffraction pattern (Cu K$_{\alpha}$ radiation) of CeRh$_2$Ga$_2$. The solid line through the data points is the two phase Rietveld refinement profile for the CaBe$_2$Ge$_2$-type primitive tetragonal structural model (space group $P4/nmm$) for CeRh$_2$Ga$_2$ and face centered cubic structural model (space group $Fd\bar{3}m$) for CeRh$_2$. The short vertical bars mark the fitted Bragg peak positions of the two phases. The lowermost curves represent the differences between the experimental and calculated intensities.}
\end{center}
\end{figure}

The powder XRD data were analyzed by the Rietveld refinement method using FullProf \cite{Rodriguez} which revealed the structure of CeRh$_2$Ga$_2$ to be CaBe$_2$Ge$_2$-type primitive tetragonal (space group $P4/nmm$) with lattice parameters $a=4.3321(1) $~\AA\ and $c= 9.7202(2)$~\AA. The  refined atomic coordinates are listed in Table~\ref{tab:XRD}. The XRD pattern (of annealed sample) and refinement profile are shown in Fig.~\ref{fig:XRD}. The refinement also showed the presence of about 5--6\%  impurity phase(s). The major impurity phase is identified to be CeRh$_2$ (about 3\%, see Fig.~\ref{fig:XRD}) which is an intermediate valence compound \cite{Barberis1982, Kappler1988,Sugawara1994}, and being paramagnetic in nature is expected to be of no consequence for the results and conclusion reached about the physical properties of CeRh$_2$Ga$_2$. The as-cast sample (XRD not shown) was found to contain about 8--10\% impurity and had very identical lattice parameters as the annealed one. All the data presented in this paper are for annealed sample.

\begin{table} 
\caption{\label{tab:XRD} Atomic coordinates obtained from the Rietveld refinement of powder XRD data of CeRh$_2$Ga$_2$ forming in CaBe$_2$Ge$_2$-type primitive tetragonal (space group $P4/nmm$) structure. Lattice parameters $a=4.3321(1) $~\AA\ and $c= 9.7202(2)$~\AA.}
\begin{ruledtabular}
\begin{tabular}{lclll}
  Atom & Wyckoff   &	 $x$ 	&	$y$	&	$z$	  \\	
   & symbol & \\
\hline
      Ce   & 2c &	 1/4 	&	1/4		&   0.2435(3)	 \\
      Rh1 & 2a &	 3/4 	&	1/4		&	0			 \\
      Rh2 & 2c &	 1/4 	&   1/4 	&	0.6128(3) 	 \\
      Ga1 & 2b &	 3/4 	&	1/4		&	1/2 		 \\
      Ga2 & 2c &	 1/4 	&   1/4 	&	0.8527(4)	 \\
\end{tabular}
\end{ruledtabular}
\end{table}

The lattice parameters of CeRh$_2$Ga$_2$ are found to be very close to that of CaBe$_2$Ge$_2$-type primitive tetragonal structured CePd$_2$Ga$_2$ \cite{Kitagawa1999}. The ratio $c/a= 2.24$ of CeRh$_2$Ga$_2$ is in the range of structurally unstable Ce-compounds  \cite{Kitagawa1999}. The value of $c/a$ is very similar to that of CePd$_2$Ga$_2$ ($c/a= 2.22$) which shows a structural transition at 125~K \cite{Kitagawa1999}. Therefore CeRh$_2$Ga$_2$ is favorably on the verge of structural instability. 

\begin{figure} 
\begin{center}
\includegraphics[width=8.5cm, keepaspectratio]{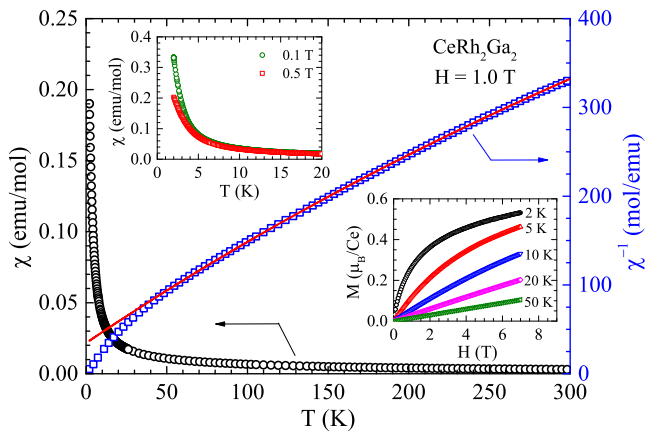}
\caption{Zero-field-cooled magnetic susceptibility $\chi$ (left ordinate) and its inverse $\chi^{-1}$ (right ordinate) of CeRh$_2$Ga$_2$ as a function of temperature $T$  for $2~{\rm K} \leq T \leq 300$~K measured in a magnetic field $H =1.0$~T\@. The solid line represents the fit by modified Curie-Weiss law for $100~{\rm K} \leq T \leq 300$~K\@. Upper Inset: Low-$T$ $\chi(T)$ for $2~{\rm K} \leq T \leq 20$~K at $H =0.1$~T and 0.5~T\@. Lower Inset: Isothermal magnetization $M$ of CeRh$_2$Ga$_2$  as a function of $H$ for $0 \leq H \leq 7$~T measured at the indicated temperatures.}
\label{fig:Chi}
\end{center}
\end{figure}

The zero-field-cooled (ZFC) dc susceptibility $\chi(T)$ and its inverse $\chi^{-1}(T)$ of CeRh$_2$Ga$_2$ measured in magnetic field $H = 1.0$~T are shown in Fig.~\ref{fig:Chi}. The $\chi(T)$ data do not show any anomaly, indicating the absence of long range magnetic ordering above 2~K\@. However, at low-$T$ the $\chi(T)$ shows a rapid increase which is more clear in the $\chi(T)$ data measured at $H =0.1$~T and 0.5~T shown in the upper inset of  Fig.~\ref{fig:Chi}. The magnitude of $\chi$ at low-$T$ is rather large. The rapid increase and high magnitude of $\chi$ very likely suggests the presence of short range correlations, and may be an indication of a possible magnetic phase transition (of ferromagnetic nature) at a temperature lower than 2~K\@. No thermal hysteresis was found between the ZFC and field-cooled (FC) $\chi(T)$ data (FC data not shown). The high-$T$ $\chi(T)$ follows the modified Curie-Weiss law, $\chi(T) = \chi_0 + C/(T-\theta_{\rm p})$. The fit of $\chi^{-1}(T)$ data with the modified Curie-Weiss law is shown in Fig.~\ref{fig:Chi}. The fit of $\chi^{-1}(T)$ over 100~K~$\leq T\leq$~300~K yielded $\chi_0 = 6.4(2)\times 10^{-4}$~emu/mol, $C=0.778(2)$~emu\,K/mol, and  $\theta_{\rm p}=-27.6(6)$~K\@. The effective moment estimated from the value of $C$ using the relation $\mu_{\rm eff} \approx  \surd (8C)\, \mu_{\rm B}$ turns out to be $\mu_{\rm eff} = 2.49(1)\, \mu_{\rm B}$/Ce which is very close to the theoretically expected value of $2.54 \, \mu_{\rm B}$/Ce for Ce$^{3+}$ ions with $J = 5/2$.

The $M(H)$ isotherms of CeRh$_2$Ga$_2$ are shown in the lower inset of Fig.~\ref{fig:Chi} for $0 \leq H \leq 7$~T at selected temperatures of 2~K, 5~K, 10~K, 20~K and 50~K\@. While the $M(H)$ isotherms at 20~K and 50~K are almost linear in $H$, the 2~K and 5~K $M(H)$ isotherms show  nonlinear behavior. The nonlinearity of $M(H)$ isotherms can be attributed to short range magnetic correlations. We also see that the 2~K isotherm tends towards a constant value, though the value of $M$ at 2~K and 7.0~T is only $M= 0.53\, \mu_{\rm B}$/Ce, which is much lower than the expected saturation value of $M_{\rm s} =g_J J= 2.14\,\mu_{\rm B}$/Ce for Ce$^{3+}$ ions (Land\'e factor $g_J=6/7$ and $J=5/2$). The low value of $M$ could be the result of crystal field effect and/or screening of moment by Kondo effect.

\begin{figure}
\begin{center}
\includegraphics[width=8cm, keepaspectratio]{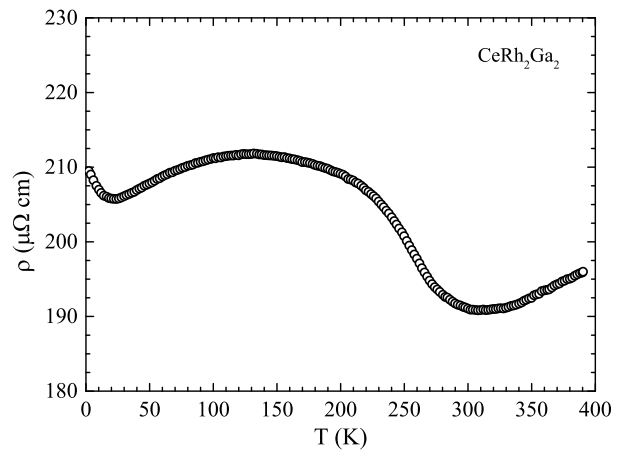}
\caption{\label{fig:Rho} Electrical resistivity $\rho$ of CeRh$_2$Ga$_2$ sample as a function of temperature $T$ for \mbox{$2~{\rm K} \leq T \leq 390$~K} measured in zero field.} 
\end{center}
\end{figure}

The $\rho(T)$ data of CeRh$_2$Ga$_2$ are shown in Fig.~\ref{fig:Rho} for $2~{\rm K} \leq T \leq 390$~K\@. The $\rho(T)$ shows the characteristic features of Kondo lattice behavior. It is seen that the $\rho$ initially decreases with decreasing $T$ as seen in metallic sample, then below about 280~K the $\rho(T)$ shows anomalous increase with a rather broad peak and a minima near 20~K below which again there is an increase in $\rho$. The anomaly near 280~K seems to be more pronounced to be accounted for by Kondo scattering alone. The heat capacity data also show a broad peak corresponding to this $\rho(T)$ anomaly. Apart from the high temperature anomaly, the overall behavior of resistivity can be understood to be the result of combined effects of Kondo screeening and crystal field. A rough estimate of Kondo temperature  $T_{\rm K}$ can be found from the value of Weiss temperature using the relation $T_{\rm K}\approx \theta_{\rm p}/4.5$ \cite{Gruner} giving $T_{\rm K} \approx 6.1$~K. The high-$T$ Kondo temperature $T_{\rm K}^{\rm h}$ can be estimated from the value of $\gamma$. Within the Coqblin-Schrieffer model  $T_{\rm K}^{\rm h}  =   {W J \pi R}/{3 \gamma}$ \cite{Yamamoto, Rajan, Tsvelick}, where $W = 0.1026 \times 4 \pi$ is the Wilson number, $R$ is the molar gas constant, and for Ce$^{3+}$ $J = 5/2$, thus $T_{\rm K}^{\rm h} \approx 216 $~K using the value of $\gamma = 130$~mJ/mol\,K$^2$ obtained below. The characteristic temperature $T_{\rm K}$ is associated with the Kondo effect on the ground state multiplet whereas the $T_{\rm K}^{\rm h}$ is associated with the Kondo effect on the excited multiplet.

\begin{figure}
\begin{center}
\includegraphics[width=8cm, keepaspectratio]{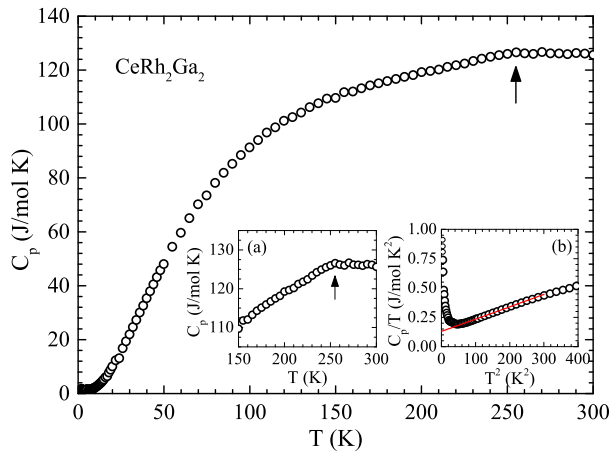}
\caption{\label{fig:HC} Heat capacity $C_{\rm p}$ of CeRh$_2$Ga$_2$ as a function of temperature $T$ for $1.85~{\rm K} \leq T \leq 300$~K measured in zero field. Inset (a): Expanded scale plot of $C_{\rm p}(T)$ showing an anomaly near 255~K. Inset (b): $C_{\rm p}/T$ versus $T^2$ plot for $1.85~{\rm K} \leq T \leq 20$~K. The solid line is a fit according to $C_{\rm p}/T = \gamma + \beta T^2 $ in $8~{\rm K} \leq T \leq 15$~K\@. The glitches near 25 K and 150 K in $C_{\rm p}(T)$ are experimental artifacts.}
\end{center}
\end{figure}

The $C_{\rm p}(T)$ data of CeRh$_2$Ga$_2$ are shown in Fig.~\ref{fig:HC} for $1.85~{\rm K} \leq T \leq 300$~K\@. Consistent with the magnetic susceptibility data the $C_{\rm p}(T)$ also does not show any evidence for magnetic phase transition. However we see a broad anomaly centered around 255~K which could be associated with the anomalous resistivity feature discussed above. The anomaly near 255~K is clearly seen in the expanded scale plot in the inset (a) of Fig.~\ref{fig:HC}. The origin of this anomaly is not clear. Since magnetic susceptibility does not show a corresponding anomaly it seems not to have a magnetic origin. We suspect that a weak structural change below 255~K could be resposible for the observed $\rho(T)$ and $C_{\rm p}(T)$ anomalies. A structural transition from tetragonal to triclinic structure has been observed in the related compound CePd$_2$Ga$_2$ (at around 125~K) as well as in LaPd$_2$Ga$_2$ (at around 62~K) \cite{Kitagawa1999}. The ratio $c/a$ of CeRh$_2$Ga$_2$ does favor the possible structural change. 

The high-$T$ $C_{\rm p}(T)$ attains a value of about 126~J/mol\,K at 300~K consistent with the limiting value according to classical Dulong-Petit law $C_{\rm V} = 3nR = 124.7$~J/mol\,K, where $R$ is the molar gas constant and $n = 5$ is the number of atoms per formula unit. At low-$T$, below 6.5~K, $C_{\rm p}(T)$ shows an upturn which is very well reflected from the $C_{\rm p}/T$ versus $T^2$ plot shown in the inset (b) of Fig.~\ref{fig:HC}. The $C_{\rm p}(T)/T$ data were fitted by $C_{\rm p}/T = \gamma + \beta T^2  $ over $8~{\rm K} \leq T \leq 15$~K which yielded the electronic coefficient $\gamma = 130(2)$~mJ/mol\,K$^2$ and lattice coefficient $\beta = 1.04(3)$~mJ/mol\,K$^4$. The obtained $\gamma$ corresponds to the density of states at the Fermi level ${\cal D}(E_{F}) = 55.1$~states/eV\,f.u.\ for both spin directions, and $\beta$ gives the Debye temperature $\Theta_{\rm D} = 211(2)$~K\@. The Sommerfeld coefficient $\gamma$ is significantly enhanced and indicates a moderate heavy fermion behavior in this compound. As $C_{\rm p}/T$ shows an increase with decreasing $T$, $C_{\rm p}/T \approx 910$~mJ/mol\,K$^2$ at 1.85~K, the $\gamma = 130(2)$~mJ/mol\,K$^2$ could be a lower bound of $T$ dependent value of $\gamma$ at low-$T$. It is also possible that the upturn in $C_{\rm p}/T$ is an indication of a transition at a temperature lower than 1.85~K\@. 

\begin{figure}
\begin{center}
\includegraphics[width=8cm]{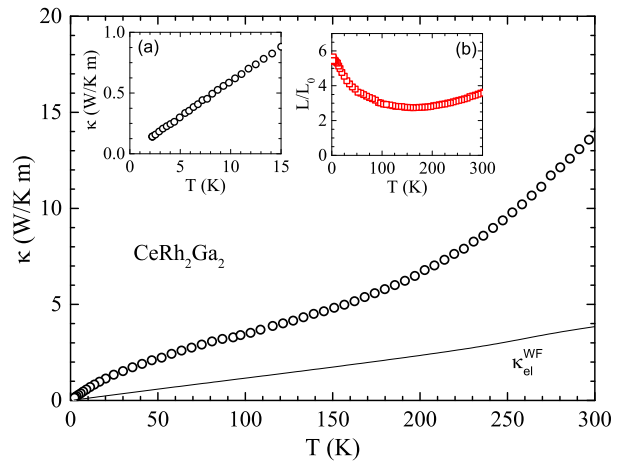}
\caption{\label{fig:TC} (Color online) Thermal conductivity $\kappa$ of CeRh$_2$Ga$_2$ as a function of temperature $T$ for $2~{\rm K} \leq T \leq 300~{\rm K}$. Solid line represents the electronic contribution to $\kappa$ estimated according to the Wiedemann-Franz law. Insets: (a) Low-$T$ $\kappa(T)$ below 15~K, and (b) $T$ dependence of reduced Lorenz number $L/L_0$.}
\end{center}
\end{figure}

The $\kappa(T)$ data of CeRh$_2$Ga$_2$ are shown in Fig.~\ref{fig:TC} for $2~{\rm K} \leq T \leq 300$~K\@. The $\kappa(T)$ does not show any evidence of a transition. As shown in the inset (a) of Fig.~\ref{fig:TC}, at low-$T$ (below 15~K) the $\kappa(T)$ shows a linear dependence on $T$. The origin of this unusual behavior of thermal conductivity is not clear. The $\kappa$ of metals and semimetals typically consists of electronic and phonon contributions.  As CeRh$_2$Ga$_2$ is metallic we estimate the electronic contribution to thermal conductivity  $\kappa_{\rm el}$ using Wiedemann-Franz law $\kappa^{\rm WF}_{\rm el} = L_0 T/\rho$ where $L_0$  is the Sommerfeld value of the Lorenz number which for metals and semimetals is $L_0 = \pi^2 k_{\rm B}^2/3e= 2.45 \times 10^{-8}~{\rm W}\,\Omega/{\rm K}^2$ ($k_{\rm B}$ is the Boltzmann constant and $e$ the electronic charge) and $\rho$ is the measured value of resitivity.  The estimated $\kappa^{\rm WF}_{\rm el}$ is shown by solid line in Fig.~\ref{fig:TC}. The $\kappa^{\rm WF}_{\rm el}$ is found to be at most 27\% of the total measured values of $\kappa(T)$ suggesting phonon dominated thermal transport in CeRh$_2$Ga$_2$. The high value of reduced Lorenz number $L(T)/L_0$ [inset (b) of Fig.~\ref{fig:TC}], where $L(T) = \kappa(T)\rho(T)/T$, also suggests the dominance of phonon contribution to thermal conductivity $\kappa_{\rm ph}$. For a system having electronic dominance in thermal transport, $L/L_0\approx 1$. In contrast, we see much larger Lorenz number in the whole range of temperature, suggesting the dominance of $\kappa_{\rm ph}$ over  $\kappa_{\rm el}$. 

\begin{figure}
\begin{center}
\includegraphics[width=8cm]{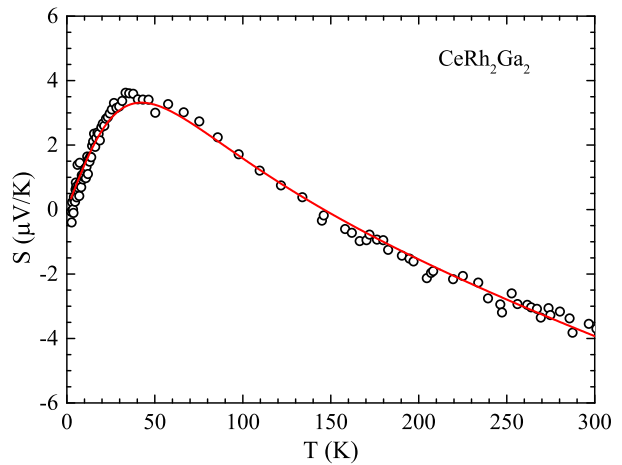}
\caption{\label{fig:Seebeck} (Color online) Seebeck coefficient $S$ of CeRh$_2$Ga$_2$ as a function of temperature $T$ for $2~{\rm K} \leq T \leq 300~{\rm K}$. Solid curve is the fit according to Eq.~(\ref{eq:S}).}
\end{center}
\end{figure}

The Seebeck coefficient $S(T)$ data of CeRh$_2$Ga$_2$ are shown in Fig.~\ref{fig:Seebeck} for $2~{\rm K} \leq T \leq 300$~K\@. With a room temperature value of $-3.5~\mu$V/K the $S$ shows a quasilinear behavior at high temperatures. The negative value of $S$ at high $T$ possibly suggests that the electrons dominate the electrical conduction. The $S$ is initially negative but shows a crossover from negative to positive value around 150~K, $S$ increases with decreasing $T$ and peaks near 35 K below which $S$ starts decreasing. A sign reversal in $S$ is usually associated with the change in the type of charge carrier or the mobility of the charge carrier. An incoherent Kondo scattering from excited CEF levels can be held responsible for the observed peak in $S(T)$ of CeRh$_2$Ga$_2$. For a three CEF level system like CeRh$_2$Ga$_2$ (tetragonal symmetry) the scattering from excited CEF level leads to a peak in $S(T)$ at a temperature $T_{\rm max} \approx \Delta_{21}/3$ where $\Delta_{\rm 21}$ is the CEF splitting energy between the ground state and the first excited state \cite{Bhattacharjee1976}. Accordingly for $T_{\rm max} = 35$~K one would expect a $\Delta_{21}$ of 105~K\@.  A weak peak at low-$T$ can also result due to the phonon-drag effect that usually occurs at a temperature $T \approx \Theta_{\rm D}/10$ to $\Theta_{\rm D}/5$. For $\Theta_{\rm D} \sim 210$~K (as estimated above) this would suggest a phonon drag peak at a temperature between 21 to 42 K, which captures the observed peak position of 35 K, though, the peak height seems to be quite stronger than what one would expect for phonon drag.
 
For a Kondo lattice system one can expect two maxima in $S(T)$, one associated with Kondo coherence (at a temperature $T_{\rm coh}$) and the other with the excited CEF levels. However, the $S(T)$ of CeRh$_2$Ga$_2$ shows only one maximum, suggesting the absence of coherence, as also inferred from the $T$ dependence of the low temperature resistivity data in Fig.~\ref{fig:Rho}. Similar behavior of $S(T)$ has been found in several Ce compounds, for example in CePdSn \cite{Adroja1992}, CeCu$_4$Ga \cite{Tolinski2010} and CeCu$_4$In \cite{Tolinski2010}. The $T$ dependence of $S$ could be described by a phenomenological resonance model of Kondo lattice heavy fermions accounting for scattering of electrons between the broad conduction band and narrow $f$-band approximated by a Lorentzian shape, given by \cite{Koterlyn2001, Koterlyn2003}
\begin{equation}
S_f(T) = \frac{2}{3}\frac{k_{\rm B}}{|e|}\pi^2\frac{T E_f}{(\pi^2/3)T^2 + E_f^2+\Gamma_f^2}
\end{equation}
where $E_f$ is the position of $f$-band (relative to Fermi energy $E_{\rm F}$) and $\Gamma_f$ is the width of the resonance peak. The $S(T)$ data were fitted by substituting $E_f = T_{\rm K}$ and $\Gamma_f = \pi \Delta_{\rm CEF}/N_f$ \cite{Koterlyn2001, Tolinski2010},  
\begin{equation}
S(T) = \frac{2}{3}\frac{k_{\rm B}}{|e|}\pi^2\frac{T T_{\rm K}}{(\pi^2/3)T^2 + T_{\rm K}^2+\pi^2 \Delta_{\rm CEF}^2/N_f^2} + aT
\label{eq:S}
\end{equation}
where $N_f = 2J+1$ is the orbital degeneracy. The Mott's diffusion term $aT$ was added to account for the the contribution other than the Kondo and CEF contributions as we do not have the $S(T)$ of nonmagnetic reference compound to separate out the $S_f(T)$ contribution. A fit of the $S(T)$ data by Eq.~(\ref{eq:S}) thus can provide an estimate of $T_{\rm K}$ and  $\Delta_{\rm CEF}$. The fit of $S(T)$ data is shown by red curve in Fig.~\ref{fig:Seebeck}.  With $N_f = 6$ for $J = 5/2$ the fit yielded $T_{\rm K} = 2.4(1)$~K, $\Delta_{\rm CEF}= 180(4)$~K and $a = 1.9\times 10^{-3}~\mu$V/K$^2$. Thus the $S(T)$ data give an estimate of overall CEF splitting energy of $\sim 180$~K in CeRh$_2$Ga$_2$. The $T_{\rm K}$ obtained this way is somewhat smaller but close to the value of $T_{\rm K} \approx 6.1$~K estimated above from $\theta_{\rm p}$. 

\section{Conclusions}

We have synthesized and investigated the physical properties of intermetallic compound CeRh$_2$Ga$_2$ by means of $\chi(T)$, $M(H)$, $C_{\rm p}(T)$, $\rho(T)$,  $\rho(H)$, $\kappa(T)$ and $S(T)$ measurements. The structural characterization using powder XRD technique revealed the room temperature structure of this compound to be CaBe$_2$Ge$_2$-type primitive tetragonal (space group $P4/nmm$). No clear evidence of a magnetic transition was observed from either of $\chi(T)$, $C_{\rm p}(T)$, $\rho(T)$ or $\kappa(T)$  measurements down to 1.8~K\@. The $\kappa(T)$ data suggest that the thermal transport is phonon dominated. The Lorenz number is found to be considerably higher than the theoretical value of $L_0$. The negative value of $S$ at high-$T$ could be a hint for an electron dominated electical conduction. The high-$T$ paramagnetic $\chi(T)$ data suggest the valence state of Ce to be Ce$^{3+}$ with an effective moment $\mu_{\rm eff} = 2.49(1)\,\mu_{\rm B}$/Ce. The $\rho(T)$ data present Kondo lattice behavior. As a consequence of Kondo scattering we find an enhanced value of Sommerfeld coefficient, $\gamma = 130(2)$~mJ/mol\,K$^2$ as inferred from the low-$T$ $C_{\rm p}(T)$ data. The enhanced value of $\gamma$ (and hence density of states) together with the Kondo lattice behavior suggests a moderate heavy fermion behavior in CeRh$_2$Ga$_2$. Both $\rho(T)$ and $C_{\rm p}(T)$ show an anomaly near 255~K\@. Further investigations are desired for understanding the origin of 255~K anomaly as well as the upturn of $C_{\rm p}/T$ at $T\leq6.5$~K\@.

\acknowledgements
DTA and VKA acknowledge financial assistance from CMPC-STFC grant number CMPC-09108. AB would like to acknowledge SA-NRF, and FRC of UJ, and ISIS-STFC for funding support. DTA would like to thank JSPS for the award of their fellowship to visit Hiroshima University.

\end{document}